# New Analytical Solutions of a Modified Black-Scholes Equation with the European Put Option

Juan Ospina
Logic and Computation Group
Quantum Econophysics Group
Mathematical Engineering Program
School of Sciences
EAFIT University
Medellín, Colombia
jospina@eafit.edu.co

**Abstract.**

Using Maple, we compute some analytical solutions of a modified Black-Scholes equation, recently proposed, in the case of the European put option. We show that the modified Black-Scholes equation with the European put option is exactly solvable in terms of associated Laguerre polynomials. We make some numerical experiments with the analytical solutions and we compare our results with the results derived from numerical experiments using the standard Black-Scholes equation.

**Keywords:** quantum econophysics, Black-Scholes equation, option pricing, special functions, kummer functions, Laguerre polynomials, symbolic computation, computer algebra, Maple.

## 1. Introduction

The standard Black-Scholes equation [1,2,3]

$$\left(\frac{\partial}{\partial t} V(t, S)\right) + r S \left(\frac{\partial}{\partial S} V(t, S)\right) + \frac{1}{2} \sigma^2 \left(\frac{\partial^2}{\partial S^2} V(t, S)\right) - r V(t, S) = 0 \qquad (0..1)$$

provides a tool to estimate the prices of different speculative financial options. According with the standard Black-Scholes equation (0.1) the price $V(t,S)$ of a speculative financial option is a function of the time $t$ and the current price $S$ of the underlying asset; and such price $V(t,S)$ depends on the volatility $\sigma$ and the risk-free interest rate $r$. From the mathematical point of view, according with the equation (0.1), the volatility $\sigma$ plays the role of a diffusivity; and the risk-free interest rate $r$ plays simultaneously the role of drift parameter and reaction rate.

In order two obtain an explicit solution of the standard Black-Scholes equation (0.1) it is necessary to specify the kind of speculative financial option that we are considering. There are two classical kinds of speculative financial options. The first one is called *European call option* and it is defined by [1]

$$V(T,S) = \begin{cases} S-K & K \leq S \\ 0 & S < K \end{cases} \quad (0.2)$$

where $K$ is named the strike price of the underlying asset and $T$ is the maturity time for the considered speculative financial option. Then, a first mathematical problem consists in to solve (0.1) with the condition (0.2).

The second classical kind of speculative financial option is called *European put option* which is defined by [1]

$$V(T,S) = \begin{cases} 0 & K \leq S \\ K-S & S < K \end{cases} \quad (0.3)$$

Then, a second mathematical problem consists in to solve (0.1) with the condition (0.3).

It is well known that the solution of (0.1) with (0.2) has the form [1]

$$V_{call}(t,S) = S\left(\frac{1}{2} + \frac{1}{2}\text{erf}\left(\frac{1}{2}\frac{\sqrt{2}\left(\ln\left(\frac{S}{K}\right) + \left(r + \frac{\sigma^2}{2}\right)(T-t)\right)}{\sigma\sqrt{T-t}}\right)\right)$$
$$- K e^{(-r(T-t))}\left(\frac{1}{2} + \frac{1}{2}\text{erf}\left(\frac{1}{2}\frac{\sqrt{2}\left(\ln\left(\frac{S}{K}\right) + \left(r - \frac{\sigma^2}{2}\right)(T-t)\right)}{\sigma\sqrt{T-t}}\right)\right) \quad (0.4)$$

and the solution of (0.1) with (0.3) is given by [1]

$$V_{put}(t,S) = -S\left(\frac{1}{2} - \frac{1}{2}\text{erf}\left(\frac{1}{2}\frac{\sqrt{2}\left(\ln\left(\frac{S}{K}\right) + \left(r + \frac{\sigma^2}{2}\right)(T-t)\right)}{\sigma\sqrt{T-t}}\right)\right)$$
$$+ K e^{(-r(T-t))}\left(\frac{1}{2} - \frac{1}{2}\text{erf}\left(\frac{1}{2}\frac{\sqrt{2}\left(\ln\left(\frac{S}{K}\right) + \left(r - \frac{\sigma^2}{2}\right)(T-t)\right)}{\sigma\sqrt{T-t}}\right)\right) \quad (0.5)$$

Using (0.4) and (0.5) it is possible to compute the theoretical prices of many speculative financial options when the relevant parameters are known.

In the speculative "real" life there are appreciable deviations of "real" prices of the speculative financial options respect to the theoretical prices prescribed by the solutions of the Black-Scholes equation. In order to close the gap between the "observed" prices of the speculative financial options and the theoretical prices, some authors have proposed two kinds of strategies: a) formulate generalized Black-Scholes equations which contain the standard Black-Scholes equation as a particular case [1]; b) formulate modified Black-Scholes equations which not necessary contain the standard Black-Scholes equation as a particular case [3].

In this work we consider the case of a modified Black-Scholes equation recently proposed [3] and we will compute certain analytical solutions in the case of the *European put option* given by (0.3). For the computation we will use computer algebra software, specifically Maple [4] and we will use some special functions of the mathematical physics such as the Kummer functions [5] and the associated Laguerre polynomials [6].

## 2. Problem

We consider here the modified Black-Scholes equation proposed recently by Y. Zheng, namely

$$\left(\frac{\partial}{\partial t} V(t,S)\right) + rS\left(\frac{\partial}{\partial S} V(t,S)\right) + \frac{1}{2}\sigma(S,t)^2\left(\frac{\partial^2}{\partial S^2} V(t,S)\right) - rV(t,S) = 0 \qquad (1)$$

When

$$\sigma(S,t) = \sigma S \qquad (2)$$

the equation (1) is reduced to the classical Black-Scholes equation

$$\left(\frac{\partial}{\partial t} V(t,S)\right) + rS\left(\frac{\partial}{\partial S} V(t,S)\right) + \frac{1}{2}\sigma^2 S^2\left(\frac{\partial^2}{\partial S^2} V(t,S)\right) - rV(t,S) = 0 \qquad (3)$$

In this paper we study two particular cases of (1): the sub-Black-Scholes case with

$$\sigma(S,t) = \sigma\sqrt{S} \qquad (4)$$

and the supra-Black-Scholes case with

$$\sigma(S,t) = \sigma S^{(3/2)} \qquad (5)$$

When (4) is replaced in (1) we obtain the sub-Black-Scholes equation with the form

$$\left(\frac{\partial}{\partial t} V(t,S)\right) + rS\left(\frac{\partial}{\partial S} V(t,S)\right) + \frac{1}{2}\sigma^2 S\left(\frac{\partial^2}{\partial S^2} V(t,S)\right) - rV(t,S) = 0 \qquad (6)$$

and when (5) is replaced in (1) we obtain the supra-Black-Scholes equation with the form

$$\left(\frac{\partial}{\partial t} V(t,S)\right) + rS\left(\frac{\partial}{\partial S} V(t,S)\right) + \frac{1}{2}\sigma^2 S^3\left(\frac{\partial^2}{\partial S^2} V(t,S)\right) - rV(t,S) = 0 \qquad (7)$$

Our problem consists is to solve respectively the equations (6) and (7) with the *European put option* given by (0.3)

## 3. Method

In this section a computational procedure will be given step by step in order to solve the problems (6)-(0.3) and (7)-(0.3). We will use Maple and we will apply some special functions of the mathematical physics such as the Kummer functions and the associated Laguerre polynomials. Specifically we will use the orthonomality property of the associated Laguerre polynomials, which has the form [7]

$$\int_0^\infty e^{(-x)} x^k \, L(n, k, x) \, L(m, k, x) \, dx = \frac{(n+k)! \, \delta_{n,m}}{n!} \tag{7A}$$

where $L(n,k,x)$ is the associated Laguerre polynomial with degree $n$ and order $k$ in the variable $x$. For the computations will be use the Maple notation

$$\text{LaguerreL}(n, k, x) = L(n, k, x) \tag{7B}$$

### 3.1. Sub-Black-Scholes equation

We look for a solution of the equation (6) with the form

$$C(x, t) = e^{(-\lambda t)} F(x) \tag{8}$$

Replacing (8) in (6) we obtain

$$\left(\frac{\partial}{\partial t}(e^{(-\lambda t)} F(S))\right) + r S \left(\frac{\partial}{\partial S}(e^{(-\lambda t)} F(S))\right) + \frac{1}{2} \sigma^2 S \left(\frac{\partial^2}{\partial S^2}(e^{(-\lambda t)} F(S))\right) - r \, e^{(-\lambda t)} F(S)$$
$$= 0 \tag{9}$$

which is reduced to

$$-\lambda \, F(S) + r S \left(\frac{d}{dS} F(S)\right) + \frac{1}{2} \sigma^2 S \left(\frac{d^2}{dS^2} F(S)\right) - r \, F(S) = 0 \tag{10}$$

According with Maple the solution of (10) is given by

$$F(S) =$$
$$e^{\left(-\frac{2rS}{\sigma^2}\right)} S \left( \text{KummerM}\left(\frac{\lambda + 2r}{r}, 2, \frac{2rS}{\sigma^2}\right) \_C1 + \_C2 \, \text{KummerU}\left(\frac{\lambda + 2r}{r}, 2, \frac{2rS}{\sigma^2}\right) \right)$$
$$\tag{11}$$

The solution (11) can be rewritten as

$$F(S) = S \, e^{\left(-\frac{2rS}{\sigma^2}\right)} \left( -\frac{\_C1 \, \text{LaguerreL}\left(\frac{-2r - \lambda}{r}, 1, \frac{2rS}{\sigma^2}\right) r}{\lambda + r} + \text{KummerU}\left(\frac{\lambda + 2r}{r}, 2, \frac{2rS}{\sigma^2}\right) \_C2 \right) \tag{12}$$

Given that we need a bounded solution when $S = 0$; and given that the function KummerU tends to $\infty$ when $S = 0$, we demand that $\_C2 = 0$; and then (12) is reduced to

$$F(S) = -\frac{S\, e^{\left(-\frac{2rS}{\sigma^2}\right)}\, \_C1\, \text{LaguerreL}\left(\frac{-2r-\lambda}{r}, 1, \frac{2rS}{\sigma^2}\right) r}{\lambda + r} \qquad (13)$$

In order to guaranty that the associated *LaguerreL* function in (13) will be a polynomial we demand that

$$\frac{-2r-\lambda}{r} = n \qquad (14)$$

where *n* is a natural number. From (14) we deduce that

$$\lambda = -n\,r - 2\,r \qquad (15)$$

Replacing (15) in (13) we get

$$F(S) = \frac{e^{\left(-\frac{2rS}{\sigma^2}\right)} S\, \text{LaguerreL}\left(n, 1, \frac{2rS}{\sigma^2}\right) \_C1}{n+1} \qquad (16)$$

Given that the sub-Black-Scholes equation (6) is a linear equation, it is possible to apply the superposition principle; and then, using (16) and (8) with (15), the general solution of (6) takes the form

$$V(t, S) = \sum_{n=0}^{\infty} \frac{S\, e^{\left(-\frac{2rS}{\sigma^2}\right)} c_n\, \text{LaguerreL}\left(n, 1, \frac{2rS}{\sigma^2}\right) e^{(r(n+2)t)}}{n+1} \qquad (17)$$

We solve (6) with the European put option

$$V(T, S) = \begin{cases} 0 & K \leq S \\ K - S & S < K \end{cases} \qquad (18)$$

and for hence, from (17) and (18) we have that

$$\begin{cases} 0 & K \leq S \\ -S + K & S < K \end{cases} = \sum_{n=0}^{\infty} \frac{S\, e^{\left(-\frac{2rS}{\sigma^2}\right)} c_n\, \text{LaguerreL}\left(n, 1, \frac{2rS}{\sigma^2}\right) e^{(r(n+2)T)}}{n+1} \qquad (19)$$

Now, multiplying the both sides of (19) by $\text{LaguerreL}\left(m, 1, \frac{2rS}{\sigma^2}\right)$ we get

$$\left(\begin{cases} 0 & K \leq S \\ -S+K & S<K \end{cases}\right) \text{LaguerreL}\left(m, 1, \frac{2rS}{\sigma^2}\right) =$$

$$\sum_{n=0}^{\infty} \frac{S\, e^{\left(-\frac{2rS}{\sigma^2}\right)} c_n \text{LaguerreL}\left(n, 1, \frac{2rS}{\sigma^2}\right) \text{LaguerreL}\left(m, 1, \frac{2rS}{\sigma^2}\right) e^{(r(n+2)T)}}{n+1} \quad (20)$$

Integrating the both sides of (20) respect to $S$ from 0 to $\infty$, the equation (20) is transformed to

$$\int_0^\infty \left(\begin{cases} 0 & K \leq S \\ -S+K & S<K \end{cases}\right) \text{LaguerreL}\left(m, 1, \frac{2rS}{\sigma^2}\right) dS =$$

$$\sum_{n=0}^{\infty} \left( \frac{c_n\, e^{(r(n+2)T)}}{n+1} \int_0^\infty S\, e^{\left(-\frac{2rS}{\sigma^2}\right)} \text{LaguerreL}\left(n, 1, \frac{2rS}{\sigma^2}\right) \text{LaguerreL}\left(m, 1, \frac{2rS}{\sigma^2}\right) dS \right) \quad (21)$$

Making the variable change given by $u = \frac{2rS}{\sigma^2}$, in the integral on the right hand side of (21) we obtain

$$\int_0^\infty \left(\begin{cases} 0 & K \leq S \\ -S+K & S<K \end{cases}\right) \text{LaguerreL}\left(m, 1, \frac{2rS}{\sigma^2}\right) dS =$$

$$\sum_{n=0}^{\infty} \left( \frac{c_n\, e^{(r(n+2)T)}}{n+1} \int_0^\infty \frac{1}{4} \frac{u\, \sigma^4\, e^{(-u)} \text{LaguerreL}(n, 1, u) \text{LaguerreL}(m, 1, u)}{r^2} du \right) \quad (22)$$

Using the orthonormalization relation of the associated Laguerre polynomials given by (7A) we derive that

$$\int_0^\infty \frac{1}{4} \frac{u\, \sigma^4\, e^{(-u)} \text{LaguerreL}(n, 1, u) \text{LaguerreL}(m, 1, u)}{r^2} du = \frac{1}{4} \frac{\sigma^4 (n+1) \delta_{n,m}}{r^2} \quad (23)$$

Replacing (23) in (22) gives that

$$\int_0^\infty \left(\begin{cases} 0 & K \leq S \\ -S+K & S<K \end{cases}\right) \text{LaguerreL}\left(m, 1, \frac{2rS}{\sigma^2}\right) dS = \sum_{n=0}^{\infty} \left( \frac{1}{4} \frac{c_n\, \sigma^4\, \delta_{n,m}\, e^{(r(n+2)T)}}{r^2} \right) \quad (24)$$

which is reduced to

$$-\frac{1}{2}\left(-2\,K\,\mathrm{hypergeom}\!\left([1,-m],[2,2],\frac{2\,r\,K}{\sigma^2}\right)r\,m^2\right.$$
$$-6\,K\,\mathrm{hypergeom}\!\left([1,-m],[2,2],\frac{2\,r\,K}{\sigma^2}\right)r\,m$$
$$-4\,K\,\mathrm{hypergeom}\!\left([1,-m],[2,2],\frac{2\,r\,K}{\sigma^2}\right)r-\mathrm{LaguerreL}\!\left(m,\frac{2\,r\,K}{\sigma^2}\right)\sigma^2\,m$$
$$-\mathrm{LaguerreL}\!\left(m,\frac{2\,r\,K}{\sigma^2}\right)\sigma^2+\mathrm{LaguerreL}\!\left(m,1,\frac{2\,r\,K}{\sigma^2}\right)\sigma^2$$
$$\left.+2\,\mathrm{LaguerreL}\!\left(m,1,\frac{2\,r\,K}{\sigma^2}\right)K\,r\right)K/(r\,(m+2))=\frac{1}{4}\frac{c_m\,\sigma^4\,\mathrm{e}^{(r(n+2)T)}}{r^2}$$

(25)

From (25) we derive that

$$c_m=-2\left(-2\,K\,\mathrm{hypergeom}\!\left([1,-m],[2,2],\frac{2\,r\,K}{\sigma^2}\right)r\,m^2\right.$$
$$-6\,K\,\mathrm{hypergeom}\!\left([1,-m],[2,2],\frac{2\,r\,K}{\sigma^2}\right)r\,m$$
$$-4\,K\,\mathrm{hypergeom}\!\left([1,-m],[2,2],\frac{2\,r\,K}{\sigma^2}\right)r-\mathrm{LaguerreL}\!\left(m,\frac{2\,r\,K}{\sigma^2}\right)\sigma^2\,m$$
$$-\mathrm{LaguerreL}\!\left(m,\frac{2\,r\,K}{\sigma^2}\right)\sigma^2+\mathrm{LaguerreL}\!\left(m,1,\frac{2\,r\,K}{\sigma^2}\right)\sigma^2$$
$$\left.+2\,\mathrm{LaguerreL}\!\left(m,1,\frac{2\,r\,K}{\sigma^2}\right)K\,r\right)K\,r\,/\,((m+2)\,\sigma^4\,\mathrm{e}^{(r(n+2)T)})$$

(26)

Replacing $m$ by $n$ in (26) and then replacing (26) in (17), we obtain

$$V(t, S) = \sum_{n=0}^{\infty} \left( -2 S \left( -2 K \text{ hypergeom}\left([1, -n], [2, 2], \frac{2 r K}{\sigma^2}\right) r n^2 \right. \right.$$

$$- 6 K \text{ hypergeom}\left([1, -n], [2, 2], \frac{2 r K}{\sigma^2}\right) r n$$

$$- 4 K \text{ hypergeom}\left([1, -n], [2, 2], \frac{2 r K}{\sigma^2}\right) r - \text{LaguerreL}\left(n, \frac{2 r K}{\sigma^2}\right) \sigma^2 n$$

$$- \text{LaguerreL}\left(n, \frac{2 r K}{\sigma^2}\right) \sigma^2 + \text{LaguerreL}\left(n, 1, \frac{2 r K}{\sigma^2}\right) \sigma^2$$

$$\left. + 2 \text{ LaguerreL}\left(n, 1, \frac{2 r K}{\sigma^2}\right) K r \right) K r \text{ LaguerreL}\left(n, 1, \frac{2 r S}{\sigma^2}\right)$$

$$\left. e^{\left(-\frac{r(2S + T\sigma^2 n + 2T\sigma^2 - t\sigma^2 n - 2t\sigma^2)}{\sigma^2}\right)} \right) / ((n+2) \sigma^4 (n+1)) \right)$$

(27)

which is the solution of (6) for the European put option (18).

### 3.2 Supra-Black-Scholes equation

We look for a solution of the equation (7) with the form (8). Replacing (8) in (7) we obtain

$$\left(\frac{\partial}{\partial t}(e^{(-\lambda t)} F(S))\right) + r S \left(\frac{\partial}{\partial S}(e^{(-\lambda t)} F(S))\right) + \frac{1}{2} \sigma^2 S^3 \left(\frac{\partial^2}{\partial S^2}(e^{(-\lambda t)} F(S))\right) - r e^{(-\lambda t)} F(S)$$
$$= 0$$

(28)

which is reduced to

$$-\lambda F(S) + r S \left(\frac{d}{dS} F(S)\right) + \frac{1}{2} \sigma^2 S^3 \left(\frac{d^2}{dS^2} F(S)\right) - r F(S) = 0 \qquad (29)$$

According with Maple the solution of (29) is given by

$$F(S) = \_C1 \text{ KummerM}\left(\frac{\lambda + r}{r}, 2, \frac{2 r}{\sigma^2 S}\right) + \_C2 \text{ KummerU}\left(\frac{\lambda + r}{r}, 2, \frac{2 r}{\sigma^2 S}\right) \qquad (30)$$

The solution (30) can be rewritten as

$$F(S) = -\frac{\_C1 \text{ LaguerreL}\left(\frac{-\lambda - r}{r}, 1, \frac{2 r}{\sigma^2 S}\right) r}{\lambda} + \_C2 \text{ KummerU}\left(\frac{\lambda + r}{r}, 2, \frac{2 r}{\sigma^2 S}\right) \qquad (31)$$

Given that we need a bounded solution when $S = 0$; and given that the function KummerU tends to $\infty$ when $S = 0$, we demand that $\_C2 = 0$; and then (31) is reduced to

$$F(S) = -\frac{\_C1 \, \text{LaguerreL}\left(\frac{-\lambda - r}{r}, 1, \frac{2r}{\sigma^2 S}\right) r}{\lambda} \tag{32}$$

In order to guaranty that the associated *LaguerreL* function in (32) will be a polynomial we demand that

$$\frac{-\lambda - r}{r} = n \tag{33}$$

where *n* is a natural number. From (33) we deduce that

$$\lambda = -n\,r - r \tag{34}$$

Replacing (34) in (32) we get

$$F(S) = \frac{\_C1 \, \text{LaguerreL}\left(n, 1, \frac{2r}{\sigma^2 S}\right)}{n + 1} \tag{35}$$

Given that the supra-Black-Scholes equation (7) is a linear equation, it is possible to apply the superposition principle; and then, using (35) and (8) with (34), the general solution of (7) takes the form

$$V(t, S) = \sum_{n=0}^{\infty} \frac{c_n \, \text{LaguerreL}\left(n, 1, \frac{2r}{\sigma^2 S}\right) e^{(r(n+1)t)}}{n + 1} \tag{36}$$

We solve (7) with the European put option (18); and for hence, from (36) and (18) we have that

$$\begin{cases} 0 & K \leq S \\ K - S & S < K \end{cases} = \sum_{n=0}^{\infty} \frac{c_n \, \text{LaguerreL}\left(n, 1, \frac{2r}{\sigma^2 S}\right) e^{(r(n+1)T)}}{n + 1} \tag{37}$$

Making the change of variable given by $u = \frac{2r}{\sigma^2 S}$, in the right hand side of (37) we get

$$\begin{cases} 0 & K \leq S \\ K - S & S < K \end{cases} = \sum_{n=0}^{\infty} \frac{c_n \, \text{LaguerreL}(n, 1, u) \, e^{(r(n+1)T)}}{n + 1} \tag{38}$$

Now, multiplying the both sides of (38) by $\text{LaguerreL}(m, 1, u) \, u^3 \, e^{(-u)}$ we obtain

$$\left(\begin{cases} 0 & K \leq S \\ K - S & S < K \end{cases}\right) \text{LaguerreL}(m, 1, u) \, u^3 \, e^{(-u)} = $$
$$\sum_{n=0}^{\infty} \frac{c_n \, \text{LaguerreL}(n, 1, u) \, \text{LaguerreL}(m, 1, u) \, u^3 \, e^{(-u)} \, e^{(r(n+1)T)}}{n + 1} \tag{39}$$

Integrating the both sides of (39) respect to *S* from 0 to ∞, the equation (39) is transformed to

$$\int_0^\infty \left( \{ \begin{matrix} 0 & K \leq S \\ K-S & S < K \end{matrix} \right) \text{LaguerreL}(m, 1, u)\, u^3\, e^{(-u)}\, dS =$$

$$\sum_{n=0}^\infty \left( \frac{c_n\, e^{(r(n+1)T)}}{n+1} \int_0^\infty \text{LaguerreL}(n, 1, u)\, \text{LaguerreL}(m, 1, u)\, u^3\, e^{(-u)}\, dS \right) \tag{40}$$

Making the change of variable given by $u = \dfrac{2\, r\, S}{\sigma^2}$, in the integral on the right hand side of (40) we obtain

$$\int_0^\infty \left( \{ \begin{matrix} 0 & K \leq S \\ K-S & S < K \end{matrix} \right) \text{LaguerreL}(m, 1, u)\, u^3\, e^{(-u)}\, dS =$$

$$\sum_{n=0}^\infty \left( \frac{c_n\, e^{(r(n+1)T)}}{n+1} \int_0^\infty \frac{2\, \text{LaguerreL}(n, 1, u)\, \text{LaguerreL}(m, 1, u)\, u\, e^{(-u)}\, r}{\sigma^2}\, du \right) \tag{41}$$

Using the orthonormalization relation of the associated Laguerre polynomials given by (7A) we derive that

$$\int_0^\infty \frac{2\, \text{LaguerreL}(n, 1, u)\, \text{LaguerreL}(m, 1, u)\, u\, e^{(-u)}\, r}{\sigma^2}\, du = \frac{2\, r\, (n+1)\, \delta_{n,m}}{\sigma^2} \tag{42}$$

Replacing (42) in (41) and using $u = \dfrac{2\, r\, S}{\sigma^2}$ we obtain

$$\int_0^\infty \frac{8 \left( \{ \begin{matrix} 0 & K \leq S \\ K-S & S < K \end{matrix} \right) \text{LaguerreL}\left(m, 1, \dfrac{2\, r}{\sigma^2\, S}\right) r^3\, e^{\left(-\dfrac{2\, r}{\sigma^2\, S}\right)}}{\sigma^6\, S^3}\, dS =$$

$$\sum_{n=0}^\infty \left( \frac{2\, c_n\, r\, \delta_{n,m}\, e^{(r(n+1)T)}}{\sigma^2} \right) \tag{43}$$

which is reduced to

$$\frac{8\,r^3}{\sigma^6}\int_0^K \frac{\mathrm{LaguerreL}\!\left(m,1,\frac{2r}{\sigma^2 S}\right)e^{\left(-\frac{2r}{\sigma^2 S}\right)}(K-S)}{S^3}\,dS = \frac{2\,c_m\,r\,e^{(r(n+1)T)}}{\sigma^2} \tag{44}$$

From (44) we derive that

$$c_n = \frac{4\,r^2}{\sigma^4\,e^{(r(n+1)T)}}\int_0^K \frac{\mathrm{LaguerreL}\!\left(n,1,\frac{2r}{\sigma^2 \Sigma}\right)e^{\left(-\frac{2r}{\sigma^2 \Sigma}\right)}(K-\Sigma)}{\Sigma^3}\,d\Sigma \tag{45}$$

Replacing $m$ by $n$ in (45) and then replacing (45) in (36), we obtain

$$V(t,S) = \sum_{n=0}^{\infty}\left(\frac{4\,r^2\,\mathrm{LaguerreL}\!\left(n,1,\frac{2r}{\sigma^2 S}\right)e^{(-r(n+1)(T-t))}}{\sigma^4(n+1)}\right.$$

$$\left.\int_0^K \frac{\mathrm{LaguerreL}\!\left(n,1,\frac{2r}{\sigma^2 \Sigma}\right)e^{\left(-\frac{2r}{\sigma^2 \Sigma}\right)}(K-\Sigma)}{\Sigma^3}\,d\Sigma\right) \tag{46}$$

which is the solution of (7) for the European put option (18).

## 4. Results

In this section, some numerical experiments with the solutions (27) and (46) will be performed with the following values for the relevant parameters

$$\sigma = 0.25,\ r = 0.03,\ K = 3,\ T = 5 \tag{47}$$

and for different values of $t$. For numerical computations the series in (27) was truncated as $\infty = 20$. The results are showed at figure 1. The red curve is for $V(0,S)$ computed with a truncation $\infty = 20$. The blue curve is for $V(3,S)$ computed with a truncation $\infty = 20$. The green curve is for $V(5,S)$ computed with a truncation $\infty = 2000$. The black line is the graphical representation of the equation (0.3). It is observed that the green curve is an excellent approximation to the black line which is a numerical verification that the analytical solution (27) satisfies the condition given by the (0.3).

The results of a second experiment with the solution (27) are showed at figure 2. Again, for numerical computations the series in (27) was truncated as $\infty = 20$. The red curve is for V(0,S) computed with a truncation ∞ = 20. The blue curve is for V(3,S) computed with a truncation ∞ = 20. The green curve is for $V_{put}(0,S)$ computed with the equation (0.5). The black curve is for $V_{put}(3,S)$ computed with the equation (0.5). The yellow line is the graphical representation of the equation (0.3). It is observed that the prices of the speculative European put option computed using (27) are lower that the prices of the speculative European put option computed using the standard Black-Scholes formula (0.5). For this reason we are naming the equation (6) and the solution (27) as sub-Black-Scholes. In correspondence with this fact, the curves derived from (27) are approaching to the equation (0.3) from below; while the curves derived from the standard Black-Scholes formula (0.5) are approaching to the equation (0.3) above.

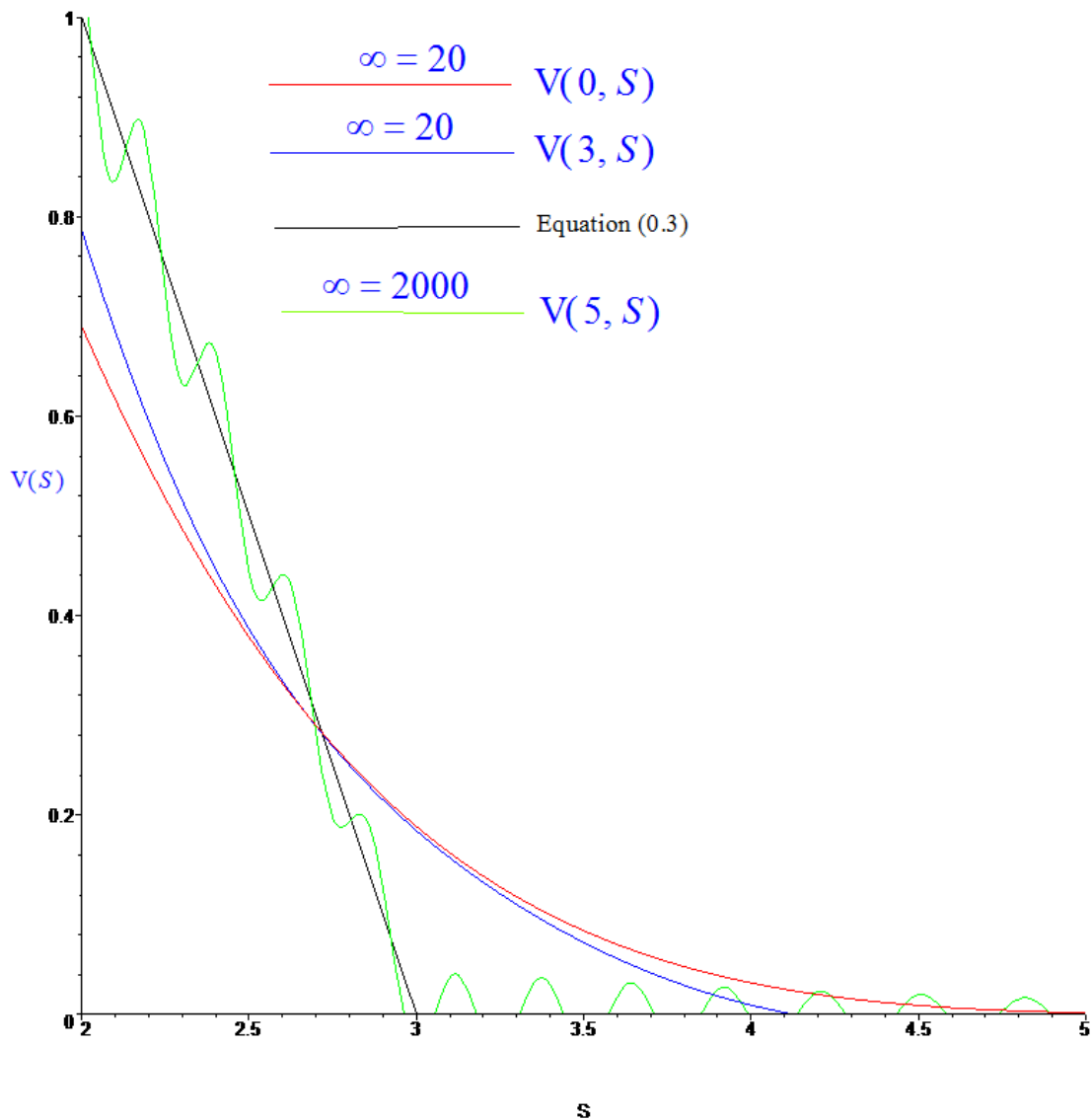

**Figure 1.** The results of a first experiment with the solution (27) with the values of parameters given by (47) and for different values of *t*. The red curve is for V(0,S)

computed with a truncation ∞ = 20. The blue curve is for V(3,S) computed with a truncation ∞ = 20. The green curve is for V(5,S) computed with a truncation ∞ = 2000. The black line is the graphical representation of the equation (0.3).

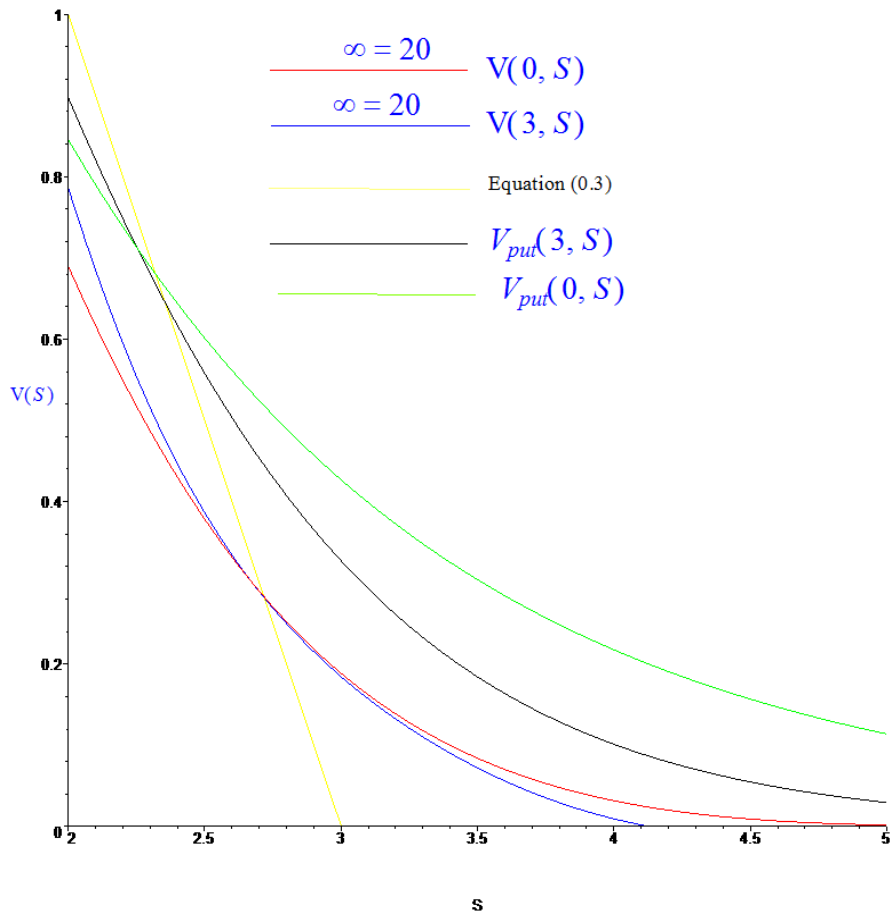

**Figure 2.** The results of a second experiment with the solution (27) with the values of parameters given by (47) and for different values of *t*. The red curve is for V(0,S) computed with a truncation ∞ = 20. The blue curve is for V(3,S) computed with a truncation ∞ = 20. The green curve is for $V_{put}$(0,S) computed with the equation (0.5). The black curve is for $V_{put}$(3,S) computed with the equation (0.5). The yellow line is the graphical representation of the equation (0.3).

Similar numerical experiments were performed with the solution (46) and the results are showed at figures 3 and 4.

In figure 3, the red curve is for V(0,S) computed with a truncation ∞ = 20. The blue curve is for V(3,S) computed with a truncation ∞ = 20. The green curve is for V(5,S) computed with a truncation ∞ = 300. The black line is the graphical representation of the equation (0.3). It is observed that the green curve is an excellent approximation to the black line which is a numerical verification that the analytical solution (46) satisfies the condition given by the (0.3).

In figure 4, The red curve is for V(0,S) computed with a truncation ∞ = 20. The blue curve is for V(3,S) computed with a truncation ∞ = 20. The green curve is for $V_{put}(0,S)$ computed with the equation (0.5). The black curve is for $V_{put}(3,S)$ computed with the equation (0.5). The yellow line is the graphical representation of the equation (0.3). It is observed that the prices of the speculative European put option computed using (46) are greater that the prices of the speculative European put option computed using the standard Black-Scholes formula (0.5). For this reason we are naming the equation (7) and the solution (46) as supra-Black-Scholes. In correspondence with this fact, the curves derived from (46) are approaching to the equation (0.3) from above.

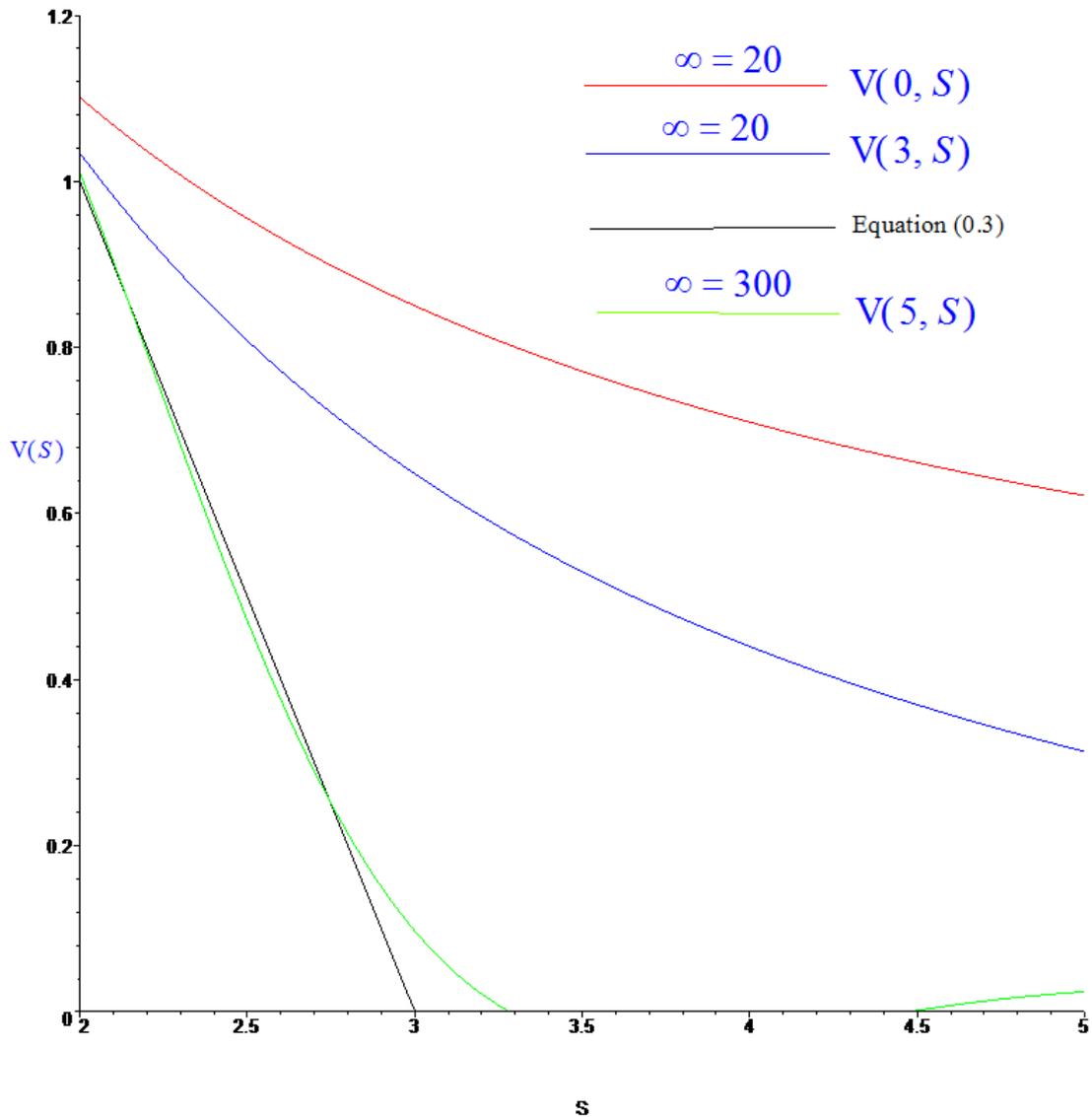

**Figure 3.** The results of a first experiment with the solution (46) with the values of parameters given by (47) and for different values of *t*. The red curve is for V(0,S) computed with a truncation ∞ = 20. The blue curve is for V(3,S) computed with a truncation ∞ = 20. The green curve is for V(5,S) computed with a truncation ∞ = 300. The black line is the graphical representation of the equation (0.3).

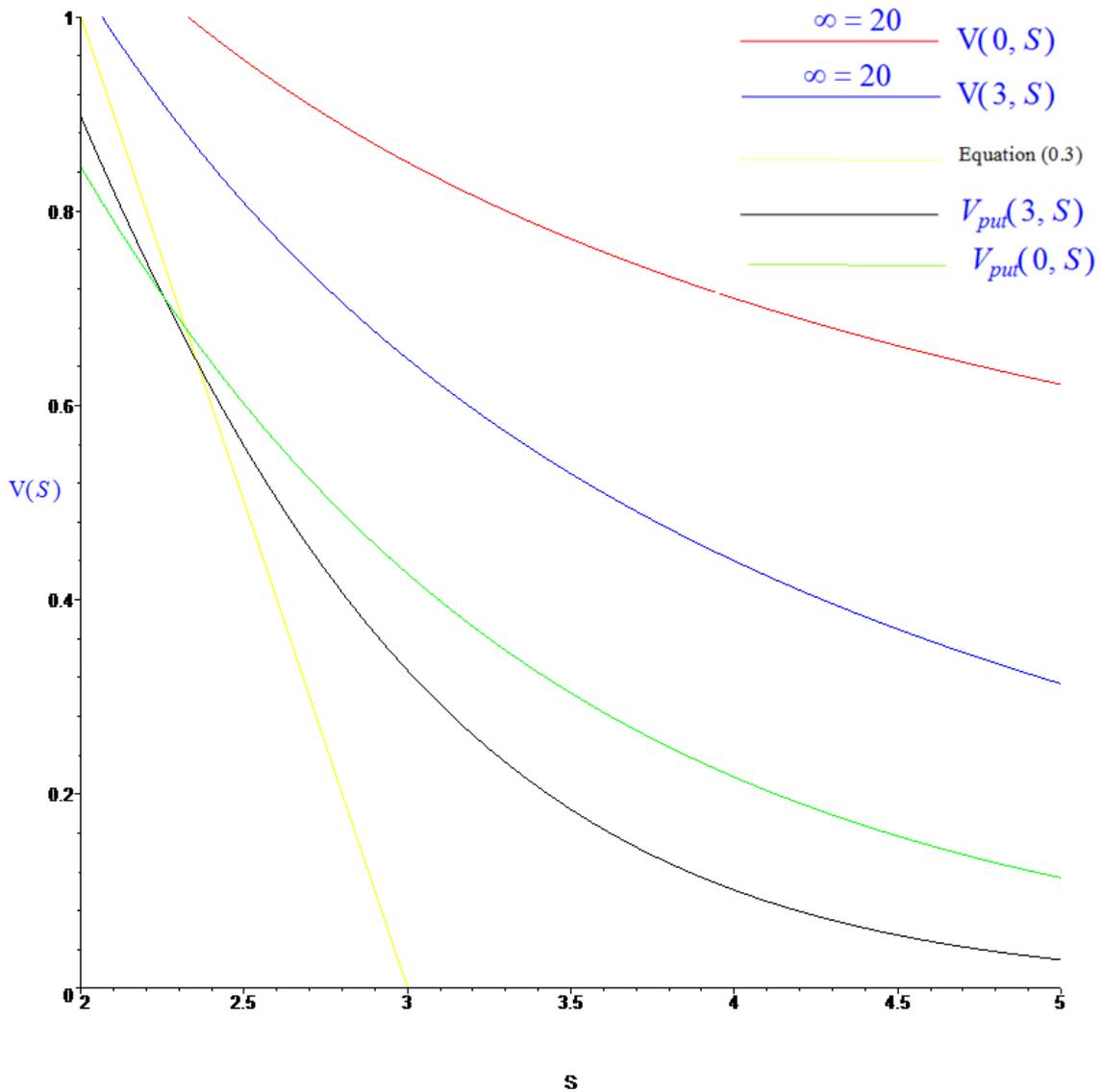

**Figure 4.** The results of a second experiment with the solution (46) with the values of parameters given by (47) and for different values of *t*. The red curve is for *V*(0,*S*) computed with a truncation ∞ = 20. The blue curve is for *V*(3,*S*) computed with a truncation ∞ = 20. The green curve is for $V_{put}$(0,*S*) computed with the equation (0.5). The black curve is for $V_{put}$(3,*S*) computed with the equation (0.5). The yellow line is the graphical representation of the equation (0.3).

## 5. Conclusions

New analytical solutions to a modified Black-Scholes equation with the European put option were found in terms of the Laguerre polynomials and Hypergeometric functions.

The Kummer functions were used as an intermediary tool. Our new solutions have been inspired from the exact solutions to certain problems in quantum mechanics such as the quantum harmonic oscillator, the hydrogen atom and the Morse potential. We claim that our new solutions will have an important role in quantum econophysics given that the standard Black-Scholes equation fails in many instances of pricing of speculative financial options. The modified Black-Scholes equations that were solved here could make better predictions than the standard Black-Scholes equation without introducing new parameters with non-direct financial meaning. As a line for future research, it is very interesting to consider the possibility to obtain a general solution for the modified Black-Scholes equation with the European put option, from which the solutions that were presented here can be considered as particular cases. Also is very interesting to attempt to solve the problem of the modified Black-Scholes equation with the European call option.

# References

1. Liviu-Adrian Cotfas, Camelia Delcea, Nicolae Cotfas, Exact solution of a generalized version of the Black-Scholes equation, arXiv:1411.2628.

2. Natanael Karjanto, Binur Yermukanova, Laila Zhexembay , Black-Scholes equation , arXiv:1504.03074

3. Y. Zheng, On Generalized Stochastic Differential Equation and Black-Scholes Dynamic Process, Proceedings of the World Congress on Engineering 2010 Vol I WCE 2010, June 30 - July 2, 2010, London, U.K.; http://www.iaeng.org/publication/WCE2010/WCE2010_pp364-367.pdf

4. Maple, www.maplesoft.com

5. Kummer functions,

https://en.wikipedia.org/wiki/Confluent_hypergeometric_function

6. Associated Laguerre Polynomials,

https://en.wikipedia.org/wiki/Laguerre_polynomials

7. Orhtonomalization for associated Laguerre polynomials,

http://mathworld.wolfram.com/AssociatedLaguerrePolynomial.html